\documentclass[12pt]{article}
\usepackage[totalheight = 23cm, totalwidth = 17cm]{geometry}
\usepackage{graphicx, subfigure}
\usepackage{hyperref}
\newcommand{\nin}{\noindent}
\newcommand{\be}{\begin{equation}}
\newcommand{\ee}{\end{equation}}
\newcommand{\bea}{\begin{eqnarray}}
\newcommand{\eea}{\end{eqnarray}}

\usepackage{amsmath}
\usepackage{amssymb}

\begin{document}

\begin{center}

{\bf{\Large The Fully Quantized Axion \\ and Dark Energy}}

\vspace{0.5cm}

D. Tanner\footnote{dtanner@ekobai.com}

King's College London, Department of Physics, WC2R 2LS, UK

\vspace{0.5cm}

{\bf Abstract}

\end{center}

\nin 
This letter reviews the exact evolution equation for the axion effective potential with the axion scale factor $f$, originally derived in \cite{Alexandre_Tanner},  and phenomenological consequences of the flat effective potential solution are discussed. It is shown that the corresponding vacuum energy ($ \sim 10^{-2}\, m_a^4$, where $m_a$ is the axion mass)  can be consistent with Dark Energy, and we compare this result to other studies relating the axion and Dark Energy.

\vspace{1cm}

{\it A Brief Review of Work on the Subject}\\  Using the widely postulated QCD axion to explain, or partially explain Dark Energy at first glance seems attractive. As one of nature's few scalar fields, it could dove-tail neatly with leading-candidate scalar field theories for Dark Energy such as quintessence.  Attempts to link the two were initiated by both the axion physics and cosmology communities shortly after the acceleration of the expansion of the universe was discovered in 1998.  The quintessence axion model \cite{Quintessence1}, (1999, 2000) uses four new pseudo scalar Goldstone bosons to create two axions, $f_q \sim \Lambda_{plank}$ and $f_a \sim 10^{12} Ge V$, to describe quintessence and the conventional CDM-QCD axion respectively.  Mass acquisition occurs at the QCD energy scale $\mu \sim \Lambda_{QCD}$ for the $f_a$ axion and at $\mu \sim 10^{-12} GeV$ for the $f_q$ quintessence axion. The latter results in an ultra light mass of $m
\sim 10^{-33} GeV$ which is equated with quintessence.  The quintaxion \cite{A Quintessential Axion}, \cite{Quintaxion} (2002, 2009) builds from the quintessence model and seeks the qualities of a very large $f$ value and a slow roll of the potential to current times. A variation of a false vaccua theory, \cite{The Cosmological Constant}, postulates that the axion field does not correspond to its true value, and this false vacuum can act as dark energy provided its lifetime is longer than the age
of the universe.  \cite{Axion Cosmology Revisited} suggests an "unstable axion quintessence" model in which the minimum of the axion potential is negative.  

\vspace{0.5cm}

{\it The Axion and Dark Energy}\\ The paper \cite{Alexandre_Tanner} from 2010 undertakes a full quantization of the axion potential, starting with an often cited form for the Euclidean bare axion potential, e.g. \cite{Axion_Kim_2010_Summary}.  Here $f$ is the scale factor and $\theta$ is, as usual, the axion phase degree of freedom.
\be\label{model} 
S_{\theta}=\int d^4x\left\lbrace
\frac{f^2}{2}\partial_\mu\theta\partial^\mu\theta +\sum_{n=1}^\infty
a_n(1 - (\cos\theta)^n)\right\}. 
\ee
In \cite{Alexandre_Tanner} treatment using a functional approach inspired by \cite{functional approach} results in an evolution equation for the effective potential $U_{eff}$ of the axion field with $f$.  The techniques used in this derivation are non-perturbative and the resulting evolution equation is exact.  The result obtained in \cite{Alexandre_Tanner} is quoted here.
\be\label{evolU} 
\dot{U}_{eff}=\frac{1}{16\pi^2f}\left[\frac{\Lambda^4}{2}-\frac{\Lambda^2}{f^2}U''_{eff}
+\frac{1}{f^4}[U''_{eff}]^2\ln\left(1+\frac{f^2\Lambda^2}{U''_{eff}}\right)
\right]. 
\ee 
Here $\dot{U}_{eff}$ is the derivative of $U_{eff}$ with respect to $f$, and $''$ represents the second derivative with respect to $\theta$.  $\Lambda$ is the high energy cut off used to regularize the theory in the quantization process. \\\\
\cite{Alexandre_Tanner} shows that in the self-interacting case, the scalar axion field is necessarily convex and flattened by spinodal instability effects when the full quantum treatment is applied.  It has been recently shown and reviewed in \cite{SSB_flat} that the effective potential of a scalar field in spontaneous symmetry breaking in the self interacting case leads to the Maxwell Construction and a flat effective potential. We also note here that while the axion field may evolve slowly over time from its inception to the present day, the time scales we are considering are microscopic fractions of a second, and we assume a time-independent $\theta$.\\\\It is noted that a valid solution of Eq.(\ref{evolU}) can indeed be a flat potential with respect to $\theta$, where $U''_{eff}=0$.
\be \label{final_result_2} \dot{U}_{eff} = \frac{\Lambda^4}{32\pi^2 f}. \ee
It is noted at this point that the variation of $f$ is used as a mathematical technique in the formulation of Eqs. (\ref{evolU}) and (\ref{final_result_2}) but it is noted that in axion physics $f$ is not a constant and indeed is usually set by hand.  It is stressed that the flattened effective potential does not depend on the axion field $\theta$ and as such $U_{eff}$ is not an expectation value of this field but rather the vacuum energy density.    A boundary condition is then assigned to Eq.(\ref{final_result_2}) such that the potential when $f = \Lambda$ is some value $U_{eff, f=\Lambda}$ leading to:
\be \label{general_solution1.1} U_{eff}(f) = \frac{\Lambda^4}{32\pi^2} \ln \left(\frac{f}{\Lambda}\right) + U_{eff, f=\Lambda}. \ee
Here $U_{eff, f=\Lambda}$ represents in integration constant solving (\ref{final_result_2}) with the boundary condition $f=\Lambda$. At the high energy level $\Lambda$, which represents the upper limit cutoff, it is considered that the evolution of the axion field is in its earliest describable form.  In this sense, for very small $\theta$, the commonly used cosine form of the axion potential (the form used in \cite{Alexandre_Tanner} prior to the quantization process) (\ref{model}) may be approximated in a polynomial expansion as in Eq. (\ref{approximation_cosine}).
\be \label{approximation_cosine} \sum_{n=1}^\infty
a_n(1 - (\cos\theta)^n) = \frac{k_1}{2}\theta^2 - \frac{k_2}{4!}\theta^4 + {\cal O}[\theta^6],
 \ee
where $k_1$ and $k_2$ are constants which are combinations of the factors $a_n$ in (\ref{model}) (note the units of $a_n$ and hence $k_1$ and $k_2$ are quartic in mass).  The axion scalar field resembles a double well potential for a scalar field $\phi$ where $\phi \equiv \phi(x)\equiv f\theta$ and we make the equivalences $\frac{k_1}{f^2}\equiv-\mu^2$ and $\frac{k_2}{f^4}\equiv-\lambda$ and $U(\phi)$ is a bare (with respect to $\phi$) potential of a double well. 
\be \label{SSB_axion} U(\phi) = -\frac{1}{2}\mu^2\phi^2 + \frac{\lambda}{4!}(\phi^2)^2. \ee
We now justify the use of the bare potential rather than an effective potential for the axion in Eq.(\ref{approximation_cosine}) and its approximation in Eq. (\ref{SSB_axion}) in the initial conditions of the solution for Eq.(\ref{general_solution1.1}).  We are at high energy levels in the realm $f \sim \Lambda$.   For such large $f$, the kinetic term in the bare action (\ref{SSB_axion}) dominates over the axion self-interactions, such that quantum fluctuations can be neglected and the effective potential is very close to the bare potential.  \\\\The approximation of the axion cosine potential Eq.(\ref{approximation_cosine}) for small $\theta$ to Eq.(\ref{SSB_axion}) has a physical significance as we may consider the axion initially as arising from a $U(1)_{PQ}$ symmetry which is spontaneously broken, a system whose classical potential for a scalar field $\phi(x)$ is the familiar double well potential as in Eq. (\ref{SSB_axion}).  It is considered that such a double well potential represents the origin of the spontaneously broken $U(1)_{PQ}$ symmetry responsible for the axion field.  An initial comment on this analysis is it does not incorporate any time parameter describing evolution of the field (the evolution is with $f$), but refers qualitatively to sequences of configurations.  The axion field emerges when the system is on the verge of spontaneous symmetry breaking.  In a typical double well scalar potential such as Eq. (\ref{SSB_axion}) this occurs when $\mu$ changes from negative (representing a symmetrical "U" shaped potential) to positive, which is a form where spontaneous symmetry breaking can occur. Thus at $\mu^2=0$, $\lambda\neq0$ (i.e. a vanishing renormalized mass squared condition) the system can be considered on the verge of spontaneous symmetry breaking as $\mu^2$ becomes $> 0$.  The significance of the result Eq. (\ref{general_solution1.1}) from \cite{Alexandre_Tanner} is that it provides an almost flat effective potential following full quantization of the bare potential in Eq (\ref{model}).  Since we have shown that this bare potential is analogous to the double well potential we seek a full quantization of Eq. (\ref{SSB_axion}).  We refer to \cite{Zee} for a computation of the one loop effective potential of this double well at $\mu=0$ in Eq. (\ref{SSB_axion}). 
\be \label{quantized SSB} V_{eff}(\phi) = \frac{1}{4!}\lambda_m \phi^4 + \frac{\lambda_m^2}{(16\pi)^2}\phi^4\left( \ln \frac{\phi^2}{m^2} - \frac{25}{6}\right) + {\cal O}[\lambda_m^3], \ee
where $m$ is an arbitrary energy scale being considered and $\lambda_m$ the energy-scale dependent effective coupling. Here $\phi$ is the classical field denoted by $\phi \equiv < \phi >$.  The term quadratic in $\lambda_m$ is the first order correction.  The flattening of the potential exhibited in Eq. (\ref{general_solution1.1}) is a tree level effect and we identify the term $U_{eff, f=\Lambda}$ with the first order correction in to the quantized potential.  Hence we equate it to the first order correction term for $V_{eff}(\phi)$ in Eq. (\ref{quantized SSB}) which is the middle term on the right hand side of Eq. (\ref{quantized SSB}) .  We acquire a negative sign for our final result for $U_{eff, f=\Lambda}$ due to the equivalences $\frac{k_1}{f^2}\equiv-\mu^2$ and $\frac{k_2}{f^4}\equiv-\lambda$ resulting in the potentials we are comparing (Eqs. (\ref{SSB_axion}) and (\ref{approximation_cosine})) having opposite signs.
\be  U_{eff, f=\Lambda} = -\frac{\lambda_m^2}{(16\pi)^2}\phi^4\left( \ln \frac{\phi^2}{m^2} - \frac{25}{6}\right). \ee
We have set a high energy limit to our theory of $\Lambda$, the regularization cut-off, and consider a constant field $\phi$ and also that $\phi, m << \Lambda$, thus satisfying our small $\theta$ approximation in Eq. (\ref{approximation_cosine}).  We further now make an assumption that $m \approx \phi$ based on the fact that both are arbitrary for the purposes of our reasoning and both are small compared to $\Lambda$.   We note we are considering the case where $\Lambda \sim f$.  We do not take the energy scale $m$ (and therefore $\phi$) as equal to $\Lambda$ as we wish to explore the behavior at energy scales lower than the cut off of our theory, $\Lambda$.  We wish to keep $m$ in our theory as a variable representing the energy scale below $\Lambda$ at which we are investigating. \\\\
In an analysis of the parameter $\lambda_m$ we have no means of further evaluating $\frac{k_2}{f^4}\equiv-\lambda$.  However, referencing other axion work, it is noted that the use of the bare potential (\ref{SSB_axion}), with $\mu = 0$, in quantization of the axion is similar to the approach taken in \cite{Bose}, where it is shown that invisible axions form a Bose-Einstein condensate.  Similarly in \cite{axion dark matter} and \cite{axion cosmology} the axion field evolution is considered in this manner.  In \cite{Bose} the authors use a $\phi^4$ scalar model for the axion and compute the effective scalar coupling constant as follows:
\be \label{scalar_coupling} \lambda = \frac{m_a^2}{f^2}\frac{m_d^3 + m_u^3}{(m_d + m_u)^3}\cong 0.35\frac{m_a^2}{f^2}, \ee
where $m_a$ is the axion mass and $m_d$ and $m_u$ are the up and down quark masses.  The authors state that this formula is obtained by using current algebra methods to derive an expression for the axion effective potential and equating the fourth-order coefficient to $\lambda$.  While $\lambda_{m}$ in (\ref{quantized SSB}) represents the scalar coupling at the energy scale $m$ compared to $\lambda$ in (\ref{scalar_coupling}) representing the coupling at the QCD energy scale, as a cursory approximation we take $\lambda \equiv \lambda_m$, which (with $f = \Lambda$) gives:
\be \label{final result 4}  U_{eff, f=\Lambda} \sim    4 \times 10^{-2}\, \frac{m^4}{f^4}m_a^4.  \ee
As our energy scale $m$ approaches the cut-off $\Lambda$ (which in our theory $\sim f$) Eq.(\ref{final result 4}) reduces to $U_{eff, f=\Lambda} \sim 10^{-2}\, m_a^4$.  Here we have assumed $\lambda << 1$.  The mass of the invisible axion has been experimentally reduced to a bound of $10^{-4}eV < m_a < 10^{-1}eV$, \cite{axion dark matter}.  \cite {Spinodal Instabilities and the Dark Energy Problem} notes a phenomenologically required energy density for dark energy in
of order $U_{DE} \sim (10^{-3} eV)^4$, which is representative of commonly quoted values.  While the result in Eq.(\ref{final result 4}) is limited in usefulness by the inconclusive nature of the ratio $\frac{m^4}{f^4} < 1$, it is interesting to note that the vacuum energy result obtained for $U_{eff}$ in Eq.(\ref{final result 4}) is of the correct order of magnitude for Dark Energy.\\

\vspace{0.5cm}

{\it Comments and Comparison with Other Models} 
\begin{itemize}

\item Quintessence theories involve a dynamical scalar field changing in space time, as evidenced by the accelerating expansion of the universe in the current era.  Research involving axions in this area was conducted by Kim and Nilles, \cite{Quintessence1}, \cite{A Quintessential Axion}.  The work focuses on linking quintessence with an ultra low mass axion whose potential has "slow rolled" down to a level associated with the required dark energy value.  The ultra low mass is obtained by considering a high $f$-valued string axion or an axion which acquires mass through contact with some hidden sector quark of ultra low mass.  In contrast, our result (\ref{final result 4}) does not rely on any non-Standard Model physics other than the proposed QCD axion.

\item The well cited works in \cite {Pseudo-Nambu-Goldstone Bosons} and \cite {Pseudo-Nambu-Goldstone BosonsII} describe a family of particles termed "pseudo-Nambu-Goldstone-bosons" (PNGB), of which the axion is an example.  These particles exhibit spontaneously broken $U(1)$ symmetry at a scale $f$ and further explicit symmetry breaking at a lower scale $\mu$, and acquiring a mass $\sim \mu^2/f$.  \cite {Pseudo-Nambu-Goldstone Bosons} treats the neutrino as a PNGB, and attempts to link its dynamical field to an effective cosmological constant for several expansion times in the universe.  We note that this approach links the mass of the neutrino-PNGB at certain eras to achieve required energy densities. \cite {Spinodal Instabilities and the Dark Energy Problem} builds on the well-cited work in \cite {Pseudo-Nambu-Goldstone Bosons} but, as in our analysis, considers spinodal instability effects on the cosine form of the neutrino's effective potential resulting in a flat energy density of $M^4$, where $M$ is the mass of a light neutrino, corresponding to a dark energy like effect. In contrast our work, in this case with QCD axion as the PNGB, results in $m_a^4$ being proportional to an energy density which we compare with Dark Energy.

\item We have characterized the early phase axion field as being flat due to spinodal instability effects.  Quintessence models require a dynamical scalar field, \cite{Copeland_Dark_Energy2}.  We have discussed only qualitatively the evolution of the axion field.  Further analysis of our result is necessary to determine how (\ref{final result 4}) could be shown to evolve into the current era.  In terms of the axion mass, $m_a$ which is a free parameter of our theory, \cite{Axion Cosmology Revisited}, for example, provides a discussion on how the axion mass may evolve with temperature scale $T$ in the early universe evolution.   Additionally we note that the parameter $\lambda$ in Eq. (\ref{scalar_coupling}) can be considered as a running coupling whose evolution with energy scale is governed by a suitable beta function.

\end{itemize}

\vspace{0.5cm}

\nin{\bf Acknowledgements} I would like to thank Jean Alexandre for guidance and comments.

\end{document}